%% file: article.tex
\documentclass{sig-alternate-05-2015} 

\pdfoutput=1
\usepackage{amsmath,amssymb}
\usepackage{graphicx}
\usepackage{subfigure}
\usepackage{acmcopyright}
\usepackage{color}
\newtheorem{defi}{Definition}

\newtheorem{theorem}{Theorem}

\begin{document}
\setcopyright{acmcopyright}

\input{macros.tex}

\markboth{L. Chen et al.}{Digital Geometry, a Survey}

\title{Digital Geometry, a Survey}
\numberofauthors{2}

\author{
\alignauthor
Li Chen\\
       \affaddr{University of the District of Columbia}\\
\alignauthor
David Coeurjolly\\
       \affaddr{Universit\'e de Lyon, CNRS, LIRIS}
}

\maketitle

\begin{abstract}
This paper provides an overview of modern digital geometry and topology through mathematical principles, algorithms, and measurements.
 It also covers recent developments in the applications of digital geometry and topology including image processing, computer vision,
 and data science.
Recent research strongly showed that digital geometry has made considerable contributions to modelings and algorithms in image segmentation, algorithmic analysis, and BigData analytics.  
\end{abstract}


\section{Introduction to Digital Geometry}

Digital geometry is the study of the geometric properties of digital and discrete objects stored in computer or electronic formats.
In general, digital geometry has two meanings: (1) Objects that are formed using digital
or integer points, which is the more narrow definition; and (2) Objects that are computerized formations of geometric data, which has
a much bigger scope.
We can view digital geometry as a subcategory of discrete geometry that has a close relationship with computerized
applications.
Digital geometry is  highly related to  computational geometry, which is more focused on algorithm design for discrete objects in Euclidean space~\cite{science-Klette2004,Che14}.

However, digital geometry has its own set of problems and challenges including those involving distance measure and the formatting of digital objects,
which are different than methods used with discrete objects. Digital geometry also has some advantages since the data can usually be sampled directly in its digital form.
For instance, contrary to discrete geometry or computational geometry, digital geometry does not require a conversion process from digital sampling to discrete forms
such as triangulation. Sampled data can be used directly.

Image processing and computer graphics are two main reasons why digital geometry was developed. First, a digital image is stored in
a digital array, and the image must be processed using some geometric properties of this type of data---digital arrays. In computer
graphics, the internal representation of the data is usually in the form of triangulation or meshes. However, when the image is displayed on
the screen, it must be digitized since the screen is a digital array.

\begin{figure}[hbt]
 \begin{center}
 \subfigure[]{\includegraphics[width=1.2in]{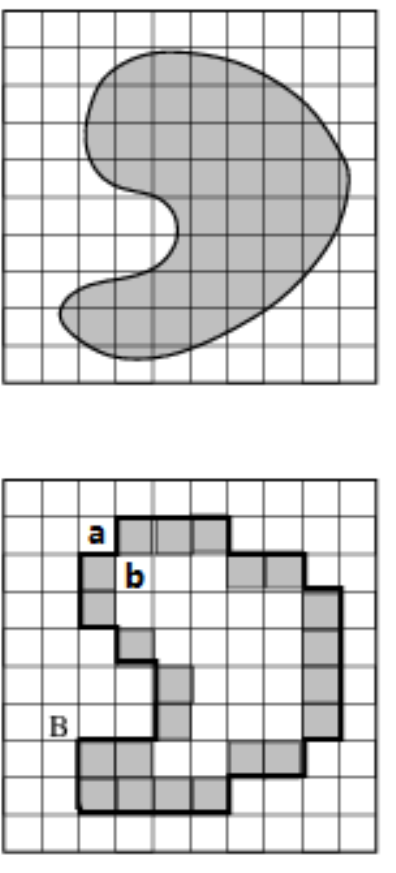}}~~~~~~~~~
 \subfigure[]{\includegraphics[width=1.2in]{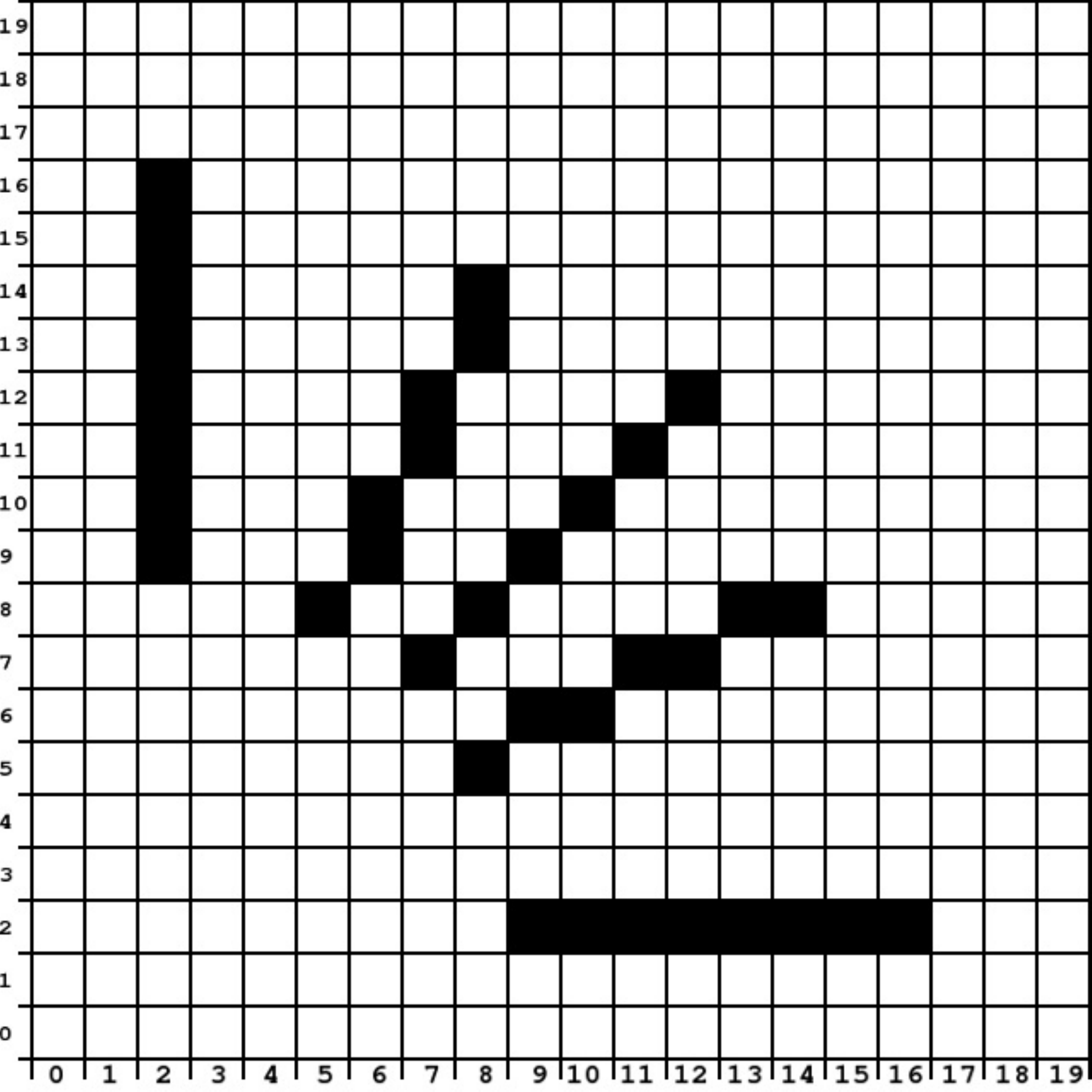}}
  \end{center}
\caption{ Digital Curves and Lines.\label{fig:f1}}
\end{figure}

Two popular examples can provide explanations as to why digital geometry is necessary. First, most image processing and computer
vision problems require extracting certain objects from the image. This process is called image segmentation, and it usually involves separating
meaningful objects from the background. Even though the boundary of an object appears to be a continuous curve, it is
actually a sequence of digital points. Identification, measurement, and extraction are all related to digital geometry. In Fig. \ref{fig:f1}(a), the top image is the original, and the bottom image is the digitized version where its boundary is a set of pixels represented as small square-blocks.

Second, in computer graphics, we usually need to draw a line or set of lines in different colors on the screen. They are drawn on 2D arrays, raster screens, instead of 2D Euclidean planes.   This is not usually a perfect line at the pixel level since it is a digital line
that requires a digital geometric algorithm such as the Bresenham algorithm to generate such a line (See Fig. \ref{fig:f1}(b)).

A problem occurs when we articulate the curves or boundary curves digitally. If we agree that a line or curve can be linked by pixels diagonally (with only one shared point at the corner of the two pixels),
then blank pixels $a$ and $b$ are also connected in Fig. \ref{fig:f1}(a). This is to say that the boundary curve of the region cannot separate a plane into two disconnected areas. Therefore, digital geometry faces a new problem that does not usually occur in continuous mathematics. This property is called the Jordan Curve theorem~\cite{Her98,science-Klette2004}.
%

Digital geometry was first considered by A. Rosenfeld~\cite{Ros70}
J. Mylopoulos, and T. Pavlidis in the early 1970s~\cite{MP71}.
In the 1980s, G. Herman and his associates studied
digital surfaces~\cite{AFH81}.
D. Morgenthaler and Rosenfeld were the first to define digital surfaces~\cite{MR81}. T. Kong and A. Roscoe conducted a
deep investigation of the properties of digital surfaces~\cite{KR85}. In the 1990s, the conference {\it Vision Geometry} drew many researchers
to this field.
Many researchers wrote monographs for this special area, see  ~\cite{Vos93,Her98,Marchand-Maillet99} to name a few.
In 2004, a comprehensive book on digital geometry was
written by  R. Klette and Rosenfeld~\cite{science-Klette2004}. There have been many remarkable developments including:
Algorithms on 3D thinning, a digital Gauss-Bonnet theorem, and
software systems for digital geometry  {\it DGtl}\cite{DGtal}. Each year, many conferences and events are held that directly relate to digital geometry.
They can be found on the official digital geometry website (\url{http://tc18.org}).

In the future, the following theoretical problems can be studied and solved: (1) Homeomorphism between two solid objects in 3D, (2) The digital Poincar\'e  conjecture and related constructive methods, and (3) Fast algorithms
for persistent analysis. Digital geometry is potentially a powerful tool that can be applied in data science and BigData, which includes topological data processing using digital
methods. Digital geometry may also prove useful in the applications of the Graphical Processing Unit (GPU). GPU is currently embedded in many computer systems as a digital array form of image processing and computer graphics.

Digital geometry usually also contains digital methods for computational differential geometry in graphics. Due to
the length limitations, this article only focuses on image related approaches. Furthermore, we do not review computerized tomography,a very sophisticated research area in digital geometry. Interested readers may refer to \cite{herman2012discrete,guedon2013mojette}.


\section{Digital Surfaces and Classification}

A digital curve can be viewed as a sequence of digital points such as the boundary of the connected region shown in Fig. 1 (a).
It can also be described mathematically as a path of vertices on a graph ~\cite{Her98,Che04}.
However, defining a digital surface is much more difficult.

\begin{figure}[hbt]
 \begin{center}
 \includegraphics[width=1.2in]{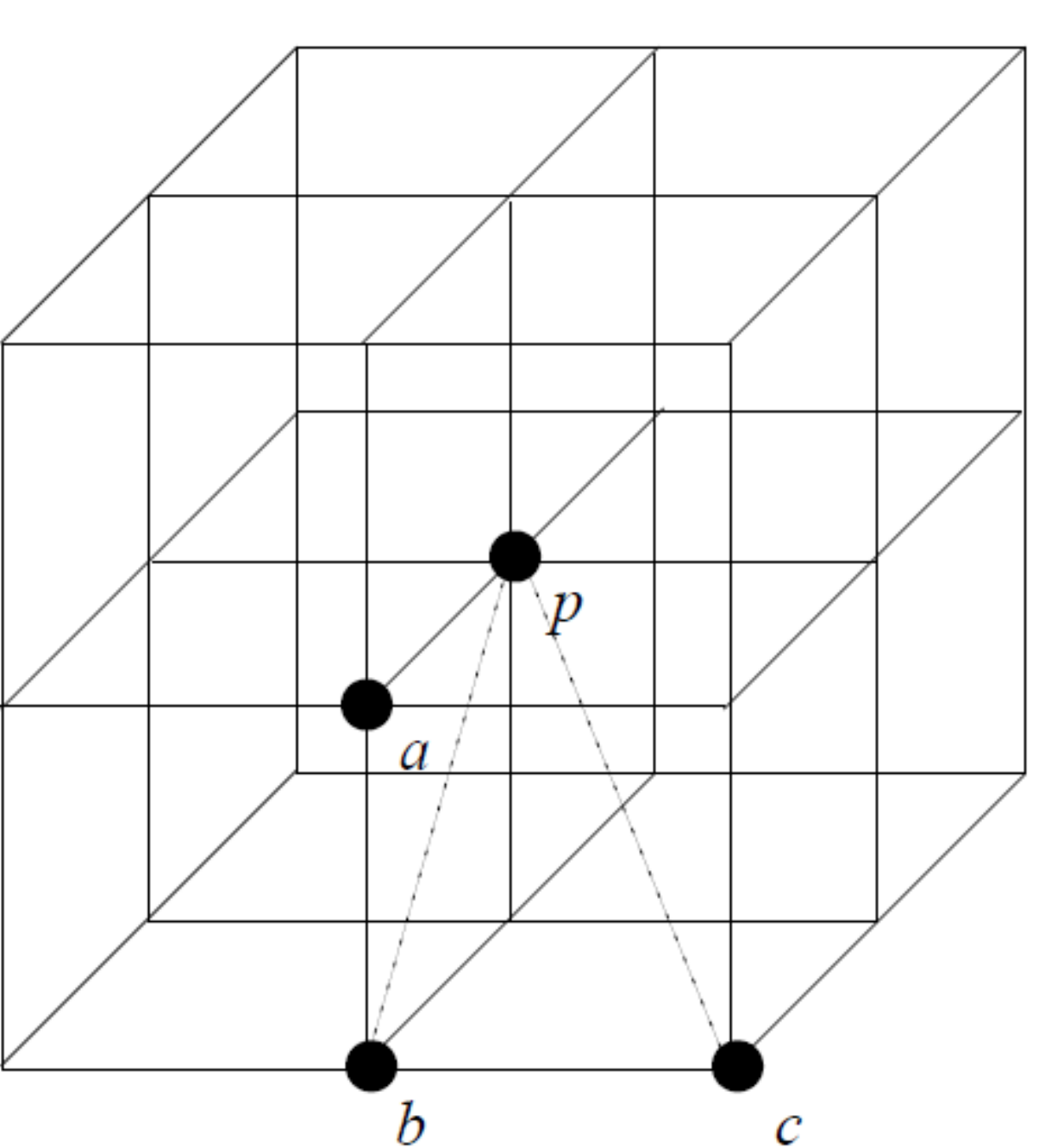}
  \end{center}
\caption{ A point $p$ and its 3D neighborhood $N_p$. There are 6 direct adjacent points such as $a$,
         18 planar diagonal adjacent points such as $b$ (including $a$'s), and 26 all adjacent points such as $c$
(including $a$'s and $b$'s).  \label{fig:f2}}
\end{figure}

A main research topic in digital geometry is how to determine the digital surface or boundary of a solid object.   In general,
we can define a digital surface based on direct adjacency and indirect adjacency~\cite{Che14}.
In 3D digital space, a digital surface is a neighborhood of a digital point $p$, denoted by $N_p$. In Fig. \ref{fig:f2},  $N_p$ contains 27 points including $p$,
meaning that  $p$ has 26 neighbors (vertices in its neighborhood).  For $p$, $a$ is a directly  adjacent point, and  $b$ and $c$ are indirectly adjacent points.
$b$ is closer to $p$ comparing the distance between $p$ and $c$.
Therefore, there are a total of 3 types of adjacencies: 6-, 18-, and 26-adjacencies shown in  Fig. \ref{fig:f2}.  A path of digital points with $\alpha$ adjacency,  $\alpha =6, 18, 26$,
is called a $\alpha$-connected path or $\alpha$-path~\cite{MR81}. Note that in 2D, there are only two adjacencies: 4- and 8-adjacency.

Digital surface was first mathematically defined by Morgenthaler and Rosenfeld (1981)\cite{MR81}. They stated that a digital surface locally
splits a neighborhood Fig. (\ref{fig:f2}) into two disconnected components where the thickness of the surface is 1.

\begin{defi}(Morgenthaler and Rosenfeld)  Let  $\alpha, \beta \in \{6,18,26\}$.  A point $p$ in
 $S$ is an $(\alpha, \beta)$-(simple) surface point, and $S(p)$ is $p$'s neighborhood in $S$ if
\protect\newline (1) $S(p)$ is an $\alpha$-component,
\protect\newline (2) $N_{p}-S(p)$ has exactly two $\beta$-components and
      $p$ is $\beta$-adjacent to both $\beta$-components,
\protect\newline (3) Each of the  $\alpha$-neighbors is $\beta$-adjacent to both $\beta$-components
of $N_{p}-S(p)$.
\end{defi}

We can see that there are a total of nine
types of digital surfaces when we choose different $\alpha’s$ and $\beta’s$.
Based on Kong and Roscoe's research, most of these cases do not exist in terms of real world examples~\cite{KR85}.
Chen and Zhang gave another definition mainly for (6,26)-surfaces, called parallel-move based surfaces~\cite{Che04}.
The advantages of this type of surface is that it can be extended to higher dimensions to define digital manifolds~\cite{Che04}.
Let $\Sigma_{m}$ be an $m$-dimensional grid space. A 0-cell is a point and 1-cell is a line-segment with two adjacent points.
A 2-cell in (6,26)-surfaces is a square with 4-points, etc. If we parallel move a 1-cell, we can get another 1-cell. In addition,
these two 1-cells form a 2-cell. We say two 2-cells are 1-adjacent if they share a 1-cell. Therefore,
we can define 1-connectedness (line-connectedness) as a path of $2$-cells where each
adjacent pair shares a $1$-cell.

\begin{defi}(Parallel-move Based)   A connected subset $S$ of $\Sigma_{m}$
 is a digital surface if any point $p\in S$ is included in some $2$-cell
of $S$, and
(1) Any two $2$-cells are 1-connected in $S$,
(2) Every 1-cell in $S$ has only one or two
        parallel-moves in $S$, and
(3) $S$ does not contain any $3$-cells.
\end{defi}

\begin{figure}
\begin{center}
 \subfigure[]{\includegraphics[width=3cm]{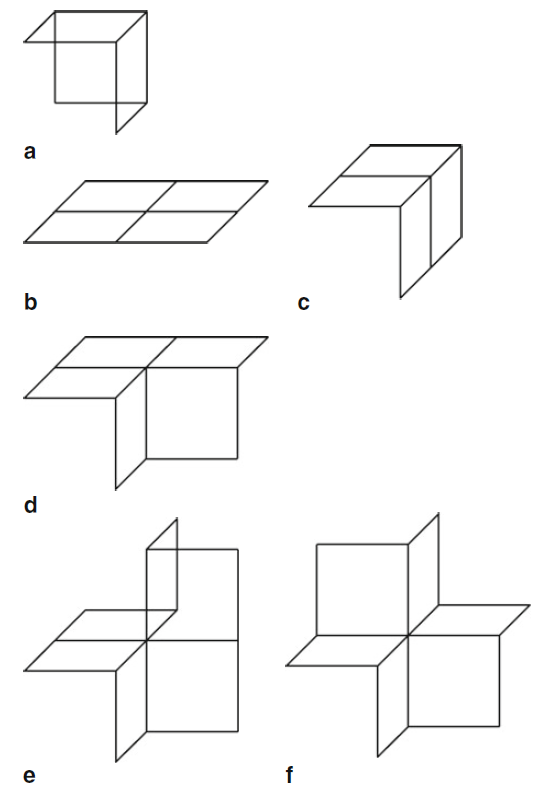}}~~~~~~~~~~~~
  \subfigure[] {\includegraphics[width=2.5cm]{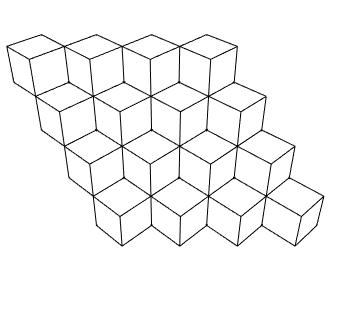}
                \includegraphics[width=2.5cm]{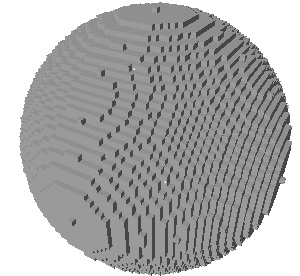}}
   \end{center}
\caption{Digital surfaces: (a) Six types of digital surfaces, (b) A digital plane and digital sphere by {\it DGtal}. \label{fig:f3}}
\end{figure}

The intuitive meaning behind this definition is that the surface is made by moving
line-segments.
Using this definition, Chen and Zhang first obtained and then proved the digital surface classification:
There are exactly 6 types of digital surface points \cite{Che04,Che14} as shown in Fig.\ref{fig:f3} (a).

Using these 6 types of digital surface points, Chen and Rong obtained
a theorem that can calculate the genus and homology groups of 3D-objects~\cite{Che14}, which we discuss in
 Section 5. There are many other definitions for digital surfaces that have been proposed by different researchers, some of which are more advanced and complex
~\cite{Che14}.


\section{Digital Topology}

In topology, a cell can be viewed as a triangle (simplex), a ball, or a cube. For instance, a disk on a plane is called a 2D-cell or 2-cell.
A cell complex is a collection of 0-cells, 1-cells, and 2-cells, etc, where a cell complex is a type of topological space.
In the 1930s, P. S. Alexandrov used an example to describe a cell complex  based on the integer grid partitions of a 2D plane~\cite{Che14}.
He used grid-cells to describe a topology. However, in combinatorial topology,  triangles or simplexes are usually used.

Even though digital topology is similar to grid-cell topology, the difference is that digital topology does not
assume we have  a set of cells pre-defining a cell complex. This is because
digital topology usually deals with 0-cells sampled by sensors without other assumptions.
Therefore, our task is not only to
find the component but also to determine a topology or cellular complex-structure  based on the collected dataset.


\subsection{Finite Topology}
In 1989, Kovalevsky proposed a method called finite topology to use in image processing\cite{Kov89}.
Kovalevsky's  method is to  encode
 images into cell complexes. The method was a nice bridge  between digital space and continuous space.

 \begin{defi}
A (abstract) cellular complex $C=(E,B,dim)$ is a set $E$ of (abstract)
elements provided with an antisymmetric, irreflexive, and transitive binary
relation $B\subset E\times E$ called the bounding relation
(or face relation) and with a dimension function
$dim: E\rightarrow \{0,1,2,...\}$ such that $dim(e_{1})<dim(e_{2})$
for all pairs $(e_{1},e_{2})\in B$.
\end{defi}

This definition simulates the
standard simplicial complex or cell
complexes.
The advantage of this definition is that it ignores continuous space and only uses
discrete objects to define the topology. However, we still need to  deal with the so called
image encoding problem:
Given a set of pixels in 2D or voxels in 3D, deciding whether
the set is a 1-complex, 2-complex, or 3-complex will give results that depend on which cells
are pre-included in $E$.

A nice property is that direct adjacency will provide a unique interpretation to encoding.
Therefore, the definition given by the parallel-move in Definition 2 becomes such an encoding.
Other digital topologies such as Khalimsky topology are also interesting to some researchers.


\subsection{Digital Manifolds}

An $n$-dimensional manifold is a topological space where each point has a neighborhood that is homeomorphic
to an $n$-dimensional Euclidean space.
How do we define $n$-dimensional digital manifolds (digital $n$-manifolds)?
In 1993, Chen and Zhang proposed a simple extension of digital surfaces to define a digital $n$-manifold\cite{Che04,Che14}.
As we can see, a digital $n$-cell can be constructed by an $(n-1)$-cell and its parallel-move. In other words,
a digital $n$-cell is a pair of two adjacent digital $(n-1)$-cells.
As we defined 1-connected above, we can define $(n-1)$-connectedness as a path of $n$-cells where each
adjacent pair shares an $(n-1)$-cell.
In general, we can define a digital $n$-manifold recursively
as follows:

\begin{defi}(Digital Manifolds)   A connected subset $M$ of $\Sigma_{m}$
 is a digital surface if any point $p\in S$ is included in some $n$-cell
of $M$, and
(1) Any two $n$-cells are $(n-1)$-connected in $M$,
(2) Every $(n-1)$-cell in $M$ has only one or two
        parallel-moves in $M$, and
(3) $M$ does not contain any $(n+1)$-cells.
\end{defi}





\begin{figure}
\begin{center}
 \subfigure[]{\includegraphics[width=2cm]{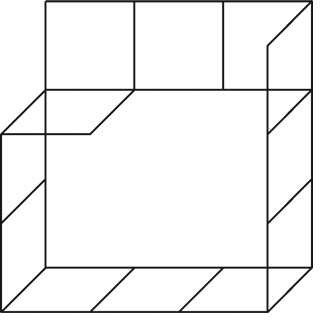}}

   \end{center}
\caption{ The Mobius band in digital space. \label{fig:f4}}
\end{figure}

 
Finding the orientability of digital surfaces or manifolds is a significant task in
topology. It is to determine if a surface or manifold contains a Mobius band.
The digital Mobius band was first discovered by  Lee and Rosenfeld (see Fig. \ref{fig:f4})~\cite{science-Klette2004}.
Chen designed an algorithm for determining whether a digital surface is orientable~\cite{Che04}.  An unsolved problem is determining the number of Mobius bands on a closed surface in high dimensional digital space. 

 A 2-cell in 3D has two normals. If a normal moves on the surface freely,
 then we want to see if these two normals can move together and be combined into one. Then, this surface is not orientable.
Let $S$ be a surface, where $N$ and $\bar N$ are
two normal lines of some surface-cell of $S$.
$S$ is nonorientable if $N$ and $\bar N$ are connected.

We can design an optimal algorithm to decide whether a
surface $S$ is orientable  or nonorientable.
In fact, we have a list of line-adjacent
(not parallel-adjacent) surface-cells of $S$.
Using the breadth-first-search on all normal-lines of $S$,
$S$ is nonorientable if the
number of connected normal-lines is greater than
 the number of 2-cells in $S$.   Otherwise,
$S$ is orientable. In summary, we have ~\cite{Che04}:

\begin{theorem} There exist an $O(|S|)$ time algorithm for the
surface orientable problem: Given a surface $S$,
decide whether $S$
is orientable.
\end{theorem}

\section{Digital Geometry Processing}
In digital geometry, image segmentation and boundary tracking were very early applications of image processing and computer vision. These problems are equivalent to extracting a component or boundary of a component in a graph. Such a graph must be a grid graph with predefined adjacencies. In 2D,
only two adjacencies can be applied: 4-adjacency and 8-adjacency. In 3D, they can be 6-, 18-, and 26-adjacency, see Fig. 2.
These algorithms have applications in CT and MRI brain images boundary tracking~\cite{BK08}. 

Besides surface tracking, extracting geometrical characteristics from object boundaries is also a very important issue. Topological modeling of digital surfaces or contours allows us to address geometric processing. For instance, we may be interested in contour length or surface area estimations, along with estimations of higher order differential quantities (normal vector fields, curvature tensors, etc.).  For these quantities, interested readers can refer to survey papers
\cite{science-Klette2004,dcoeurjo_ChapEstimateur}.
We first describe the mathematical framework used to validate a given estimator.

\subsection{Digital Labeling and Region Growing Methods for Image Segmentation}

An approach suggested by Rosenfeld was to find a threshold for an image and then divide the image into foreground and background.
Assigning the value 1 to foreground pixels and 0 to background pixels. Afterwards, we can use transitive closure to
find a connected component of 1's. This component is a segment of the image, and the  algorithm is
called the labeling algorithm.  Pavlides realized that one can use the well-known
depth-first-search or breadth-first-search methods to find the connected components~\cite{Pav82}.
This approach is similar to the region-growing method: Starting at a seed pixel, then search for all surrounding pixels with the value of 1 until reaches the 0-valued pixels.

\sloppy The region-growing method is a topological algorithm that is based on the Jordan's curve theorem stating that
a closed curve separates a plane into two disconnected components.
This is because, to use region-growing, one must assume 4-connectivity for foreground pixels since 8-connectivity does not
have such a Jordan property.
We can see here that the topology, algorithms, and image processing have been combined perfectly (Fig. ~\ref{fig:herman} (a)).

The recent work shows the great deals of interests of this from mathematical theory~\cite{ciesielski2016general,xian2016neutro} to applications, not even in image processing but also in anthropology~\cite{spradley2017smooth} and trafic control related to BigData~\cite{an2018network}.    







\subsection{3D Tracking and Medical Imaging}

\begin{figure}
\begin{center}
\subfigure[] {\includegraphics[width=4.5cm]{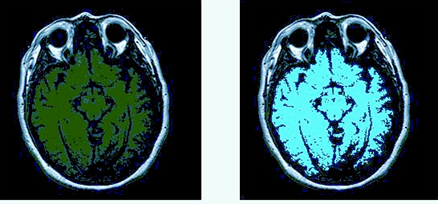}}~~
 \subfigure[]{\includegraphics[width=2.0cm]{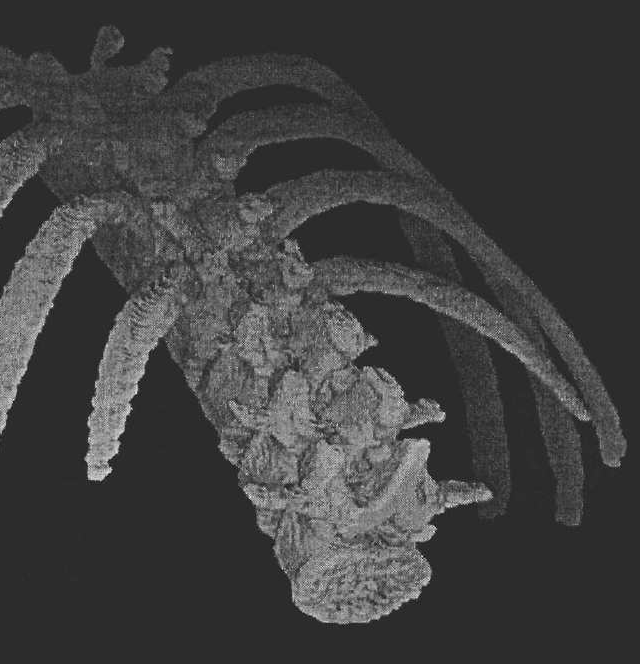}}~~
   \end{center}
\caption{Real image applications: (a)Segmentation using region-growing, and (b) Boundary tracking provided by Herman. \label{fig:herman}}
\end{figure}

Herman {\it et al} developed the so-called the algorithm of fat-flies for 3D surface tracking \cite{Her98}.  The algorithm
is a parallelized extension of the chain-code algorithm in 2D. In the algorithm, when a boundary 2-cell is found by a fat, it is marked and the fat will be cloned into 4 fats, each going in a different direction (along with
the 4-edges of this 2-cell). 
The recursive process is performed until all boundary cells are marked. The algorithm could be especially fast
for current BigData mechanisms using cloud computing (see Fig. ~\ref{fig:herman} (b)).



\subsection{Multigrid Convergence Analysis}
\label{sec:multi}
Multigrid convergence analysis is a convenient tool for evaluating the efficiency of a differential estimator. For instance, we design an estimator on a 1D digital curve. We want the estimated quantity to be close to the Euclidean quantity on a continuous curve if the digital curve corresponds to its digitization. Furthermore, the error should approach zero when the grid-step used in the digitization process approaches zero. More formally, the multigrid convergence framework considers a digitization process parameterized by a grid-step. Usually, Gauss digitization is considered. For a given family of compact subsets,  $\Shapes$ of $\R^d$, the Gauss digitization in $\Shape\in\Shapes$ is defined as follows:
\begin{equation}
  \label{eq:gauss}
  \DigF{\Shape}{h} \EqDef \left\{ \vz \in \Z^d, (h\cdot\vz)\in \Shape \right\}\,,
\end{equation}
where $h\cdot\vz$ is the uniform scaling of $\vz$ by factor ${h}$.
%
%
From this multigrid framework, the multigrid convergence of an estimator can be defined as follows:
\begin{defi}[Multigrid convergence]
  \label{def:multigrid-convergence}
  Given a digitization process, $\AnyDig$, a digital geometric
  estimator $\hat{E}$ of some global geometric quantity $E$ is {\em multigrid
    convergent} for the family of shapes {\Shapes} if and only if, for any $\Shape
  \in \Shapes$, there exists a grid step $h_\Shape>0$ such that
  \begin{equation}\forall 0< h < h_\Shape,\quad
  |\hat{E}(\AnyDigF{\Shape}{h},h) - E(\Shape) | \le \tau_{\Shape}(h),
  \end{equation}
  where $\tau_{\Shape}: \R^{+} \setminus\{0\} \rightarrow \R^+$ has null limit at
  $0$. This function defines the speed of convergence of $\hat{E}$
  towards  $E$.
\end{defi}
This definition only considers global quantities such as length or surface area. Similar definitions for local quantities on boundaries $\partial\Shape$ (normal vector, curvature) exist, which require formalizing the mapping between $\partial\Shape$ and the topological boundary of $\DigF{\Shape}{h}$ (see
\cite{DBLP:journals/jmiv/LachaudT16}).

\begin{figure}
\begin{center}
\includegraphics[width=3.0cm]{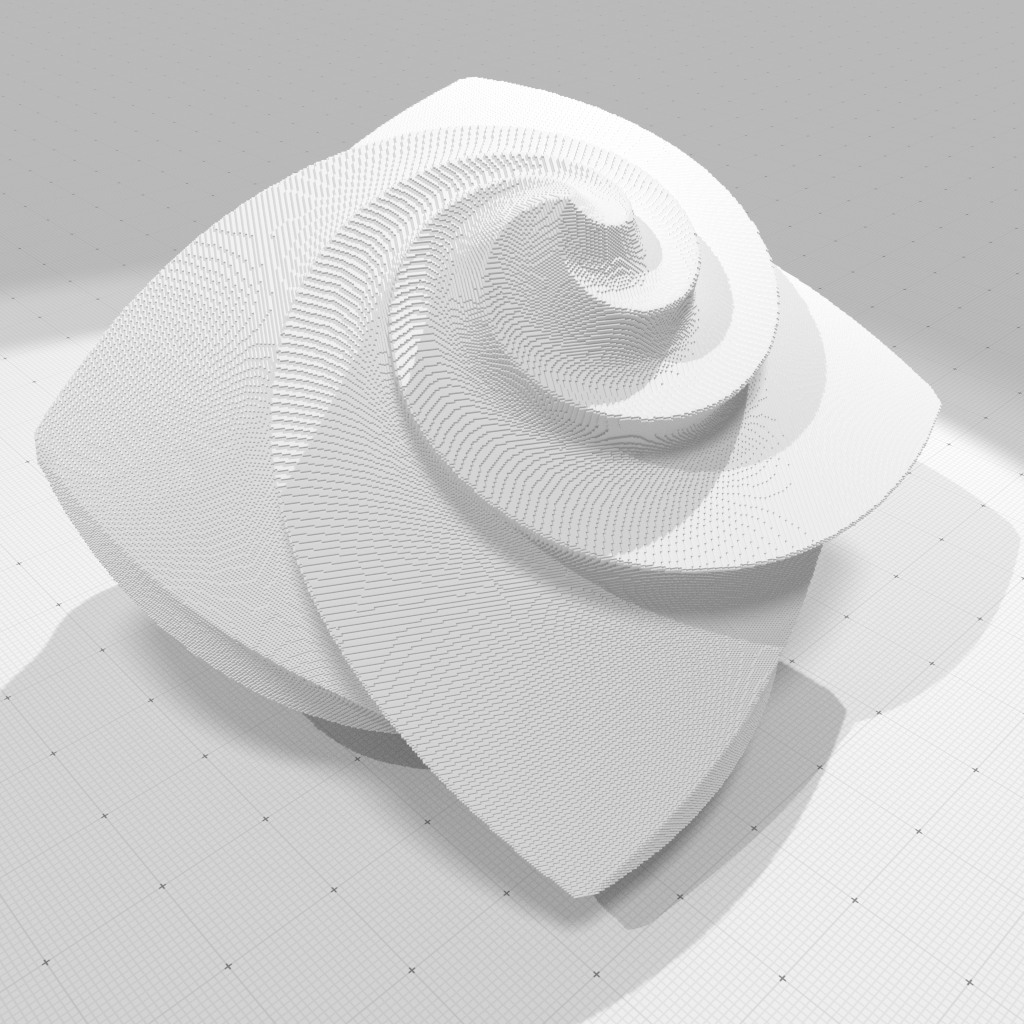}
\includegraphics[width=3.0cm]{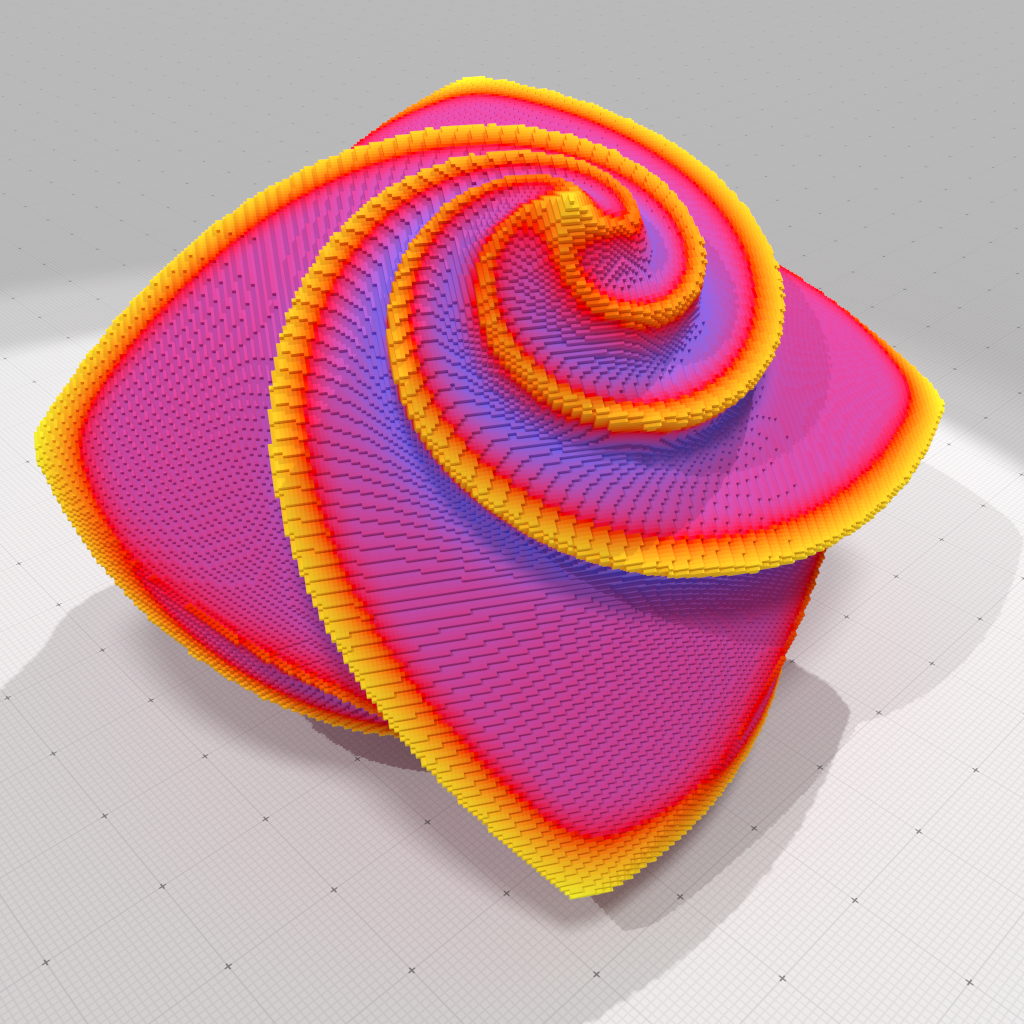}
\includegraphics[width=0.2cm,height=2cm]{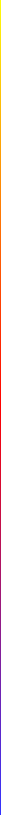}
\includegraphics[width=3.0cm]{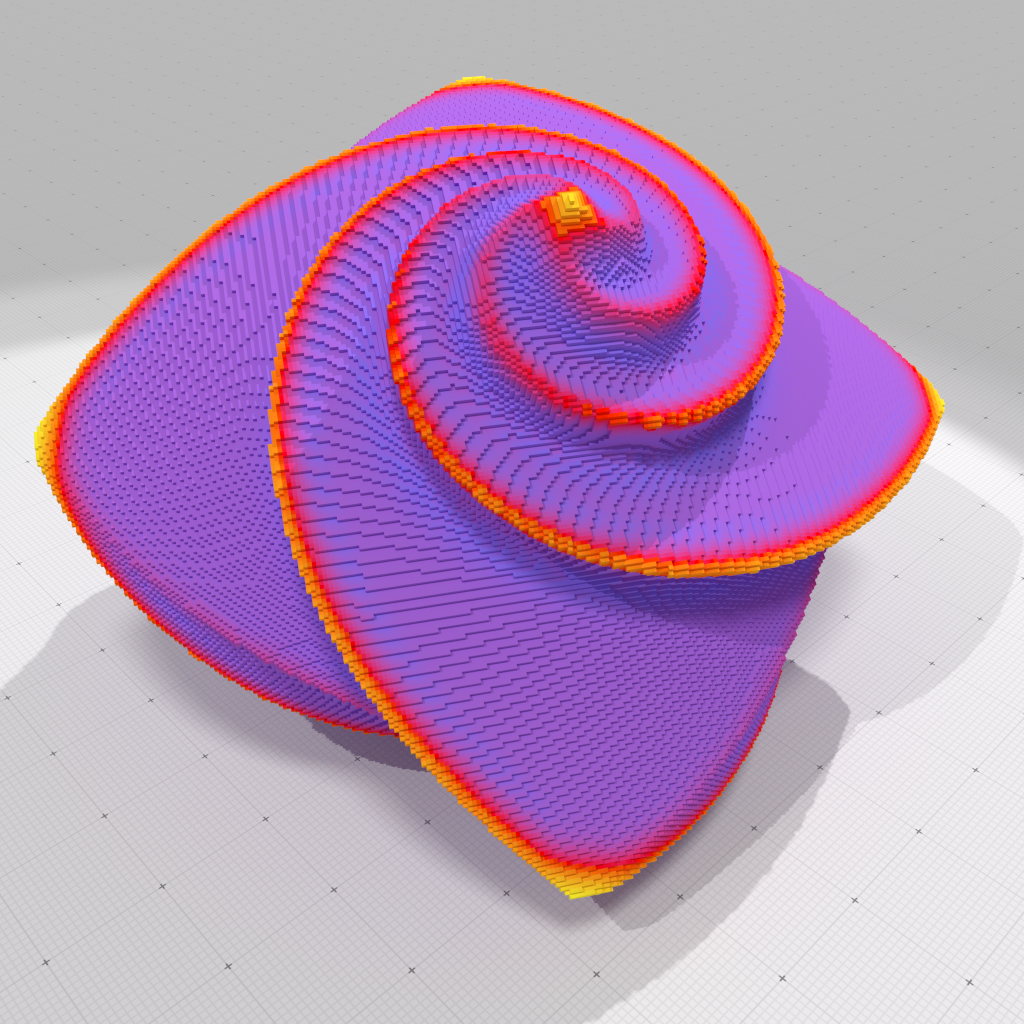}
\includegraphics[width=3.0cm]{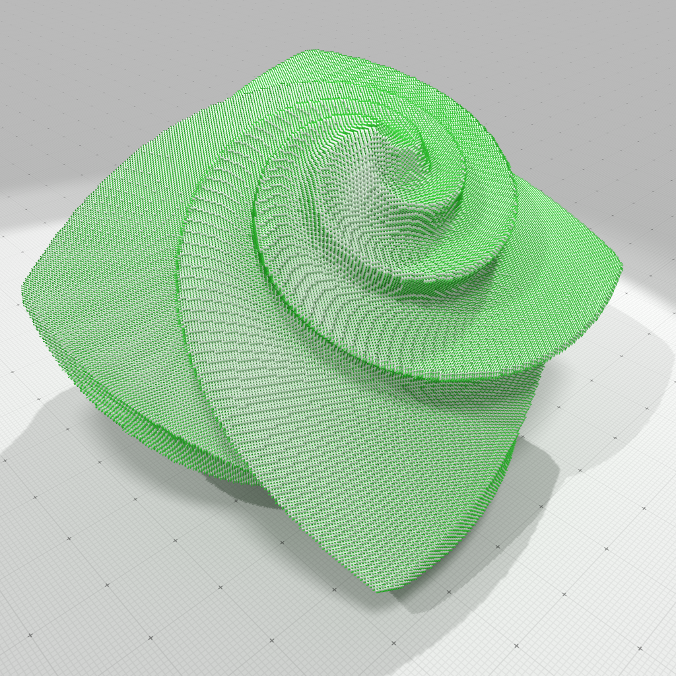}
\includegraphics[width=3.0cm]{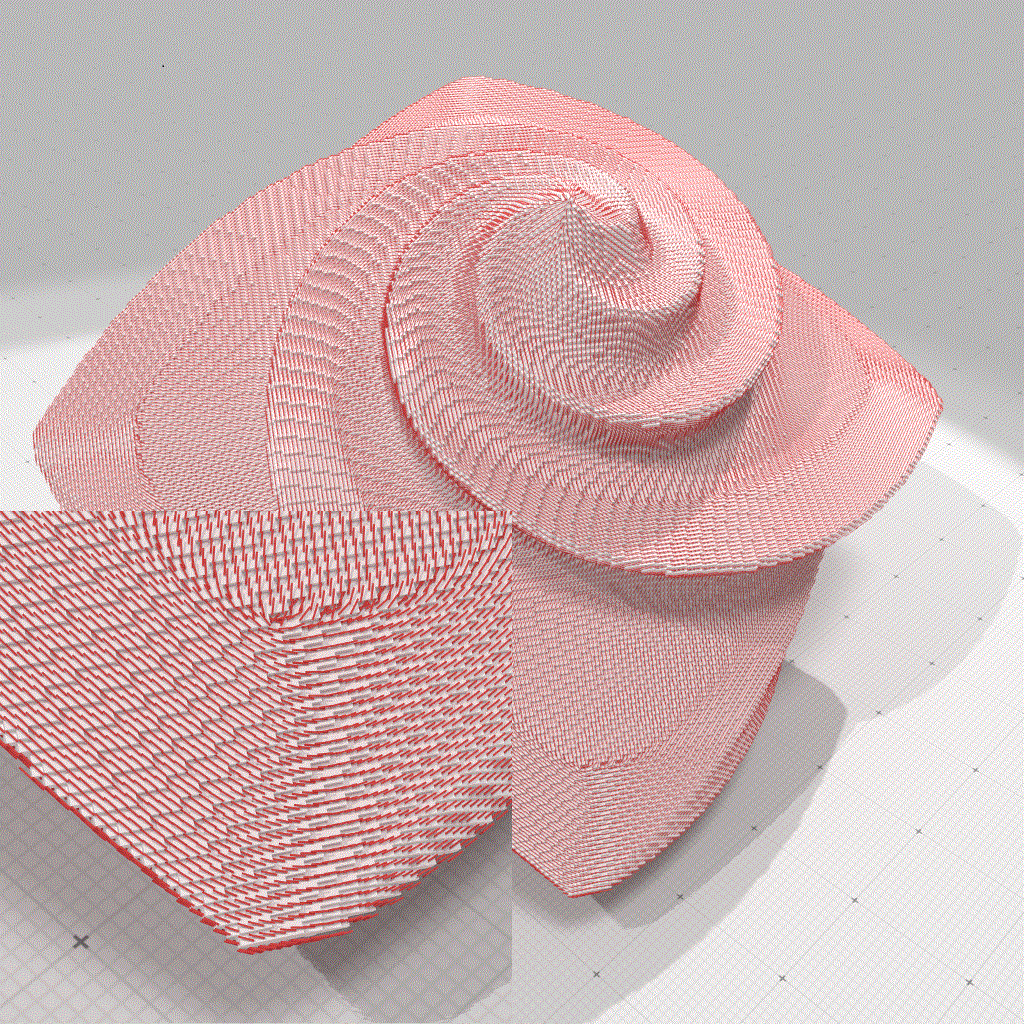}
\includegraphics[width=3.0cm]{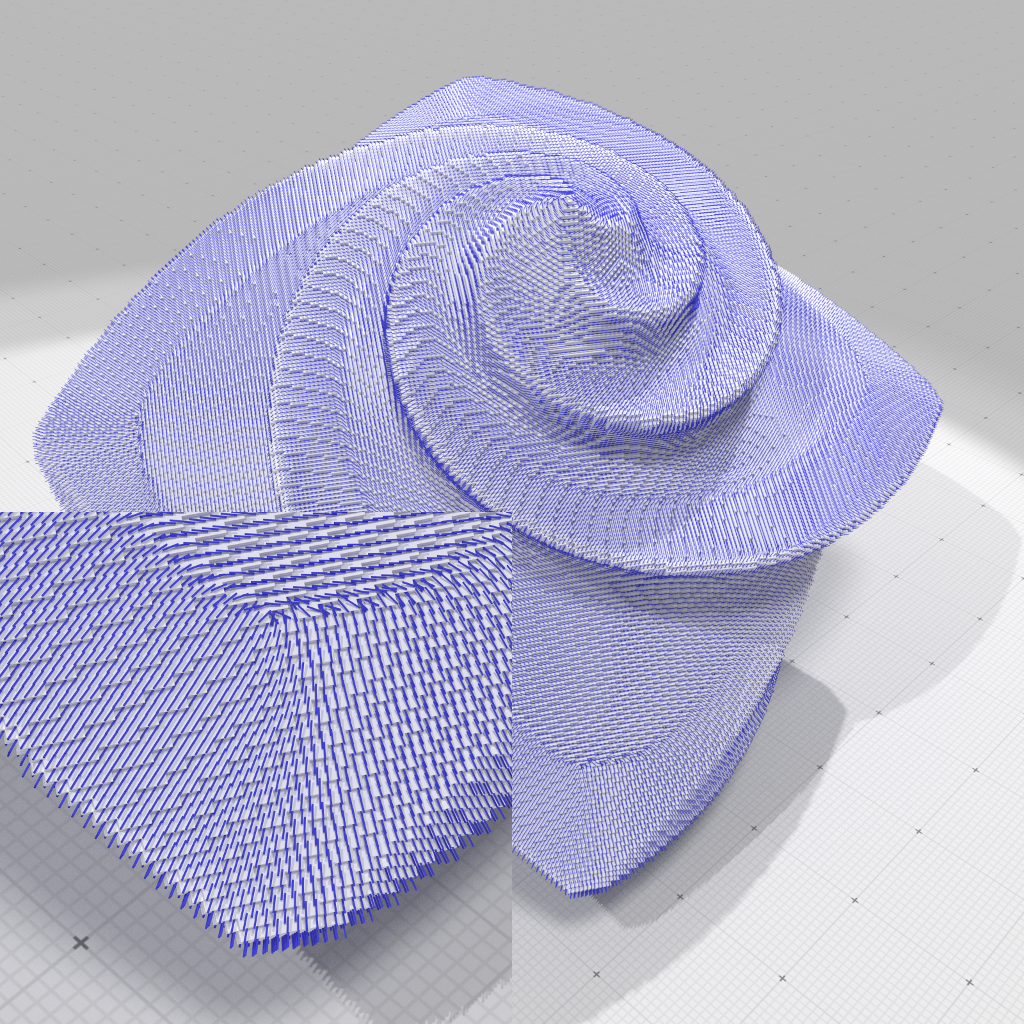}
\end{center}
\caption{Digital integral invariant based approach to estimating the mean curvature, Gaussian curvature, normal vector field, first principal curvature direction, and second principal curvature direction on a shape~\cite{DBLP:journals/cviu/CoeurjollyLL14}.\label{fig:curvature}}
\end{figure}

As mentioned above, several algorithms have multigrid convergence proofs to estimate the length, surface area, normal vectors, and curvature tensors on digital surfaces. In Figure~\ref{fig:curvature}, we illustrate the curvature tensor estimation on 3D digital surfaces.
Besides the theoretical properties, digital estimators are also compared in terms of robustness to noise and computational efficiency. For high order local differential estimators, most approaches consider a parameter specifying the size of a neighborhood or kernel radius around a given surface point. In a theoretical analysis, such a radius is parameterized by the grid-step $h$ to obtain multigrid convergence (see for example \cite{DBLP:journals/cviu/CoeurjollyLL14}). However, this parameter may be difficult to set for a given digital shape that is not defined by the multigrid digitization approach. Parameter free estimators exist for 1D digital curves in terms of length, tangent, and curvature estimation~\cite{DBLP:journals/jmiv/LachaudT16}.
Parameter free estimators on digital surfaces remain open problems.

On 1D digital contours, many estimators use simple geometric objects such as digital straight segments and digital circular arcs. Furthermore, many convergence proofs are consequences of the geometric and arithmetic properties of these elementary objects.



\subsection{Volumetric Analysis of Digital Shapes}
\label{seq:volumetric}
\begin{figure}
\begin{center}
\subfigure[]{\includegraphics[width=2.7cm]{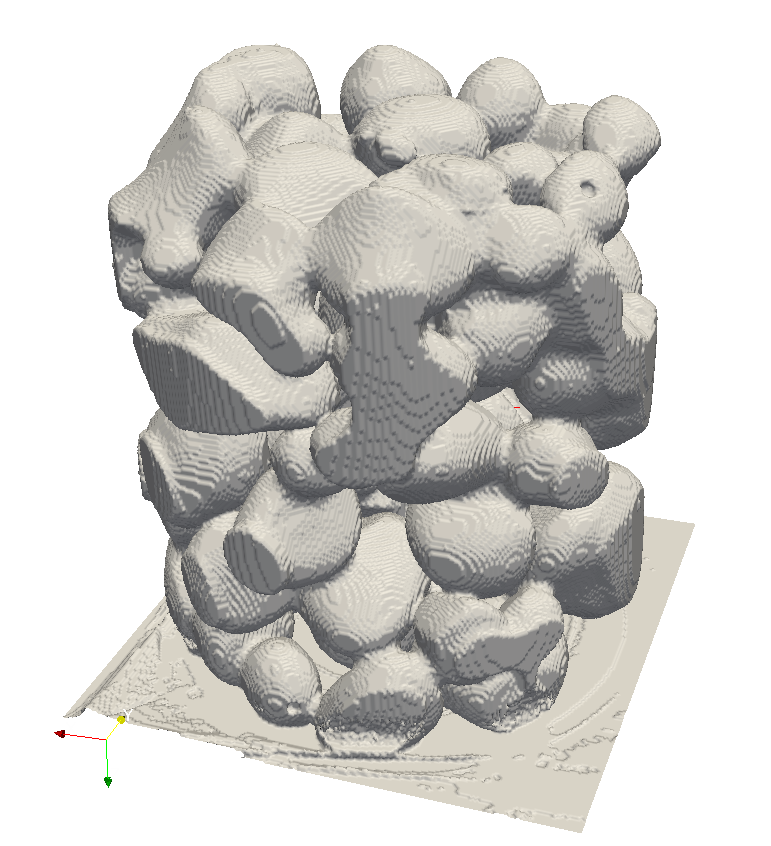}}
\subfigure[]{\includegraphics[width=2.7cm]{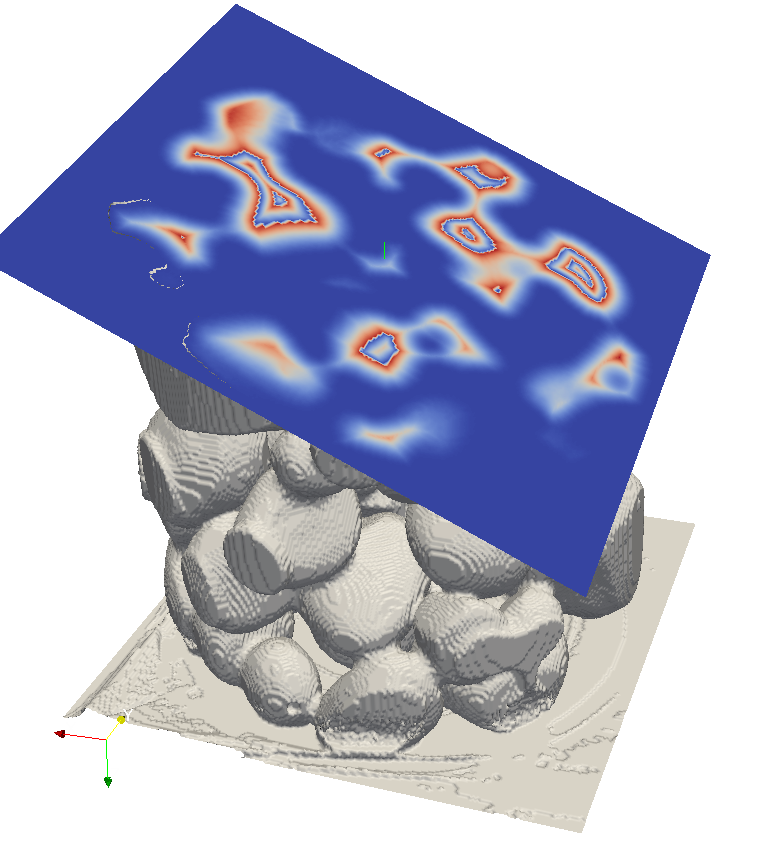}}
\subfigure[]{\includegraphics[width=2.7cm]{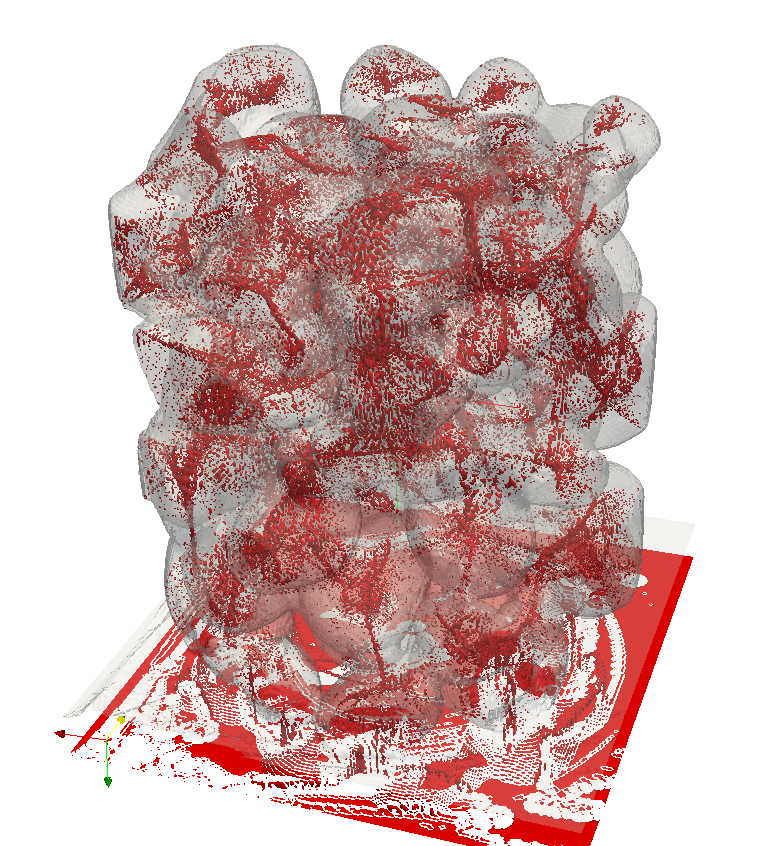}}
\end{center}
\caption{Analysis of a volumetric microtomographic snow sample in 3-D: $(a)$ The input
  object, $(b)$ Its distance transformation, $(c)$ Its discrete medial
  axis (the centers of maximal balls are in red) and a mean curvature
  labeling of its boundary.\label{fig:DT2D3D}}
\end{figure}

Metric based descriptions and analyses of shapes are fundamental notions in image analysis
and processing. In classical tools, the distance transform
(DT)
~\cite{fabbri} of a binary image
$I: \mathbb{Z}^d\rightarrow \{0,1\}$ consists of labeling
each point of a digital object $E$, defined as pixels with a value of 1 under $I$,
with its shortest distance to the complement of $E$ (see Fig. \ref{fig:DT2D3D}-$b$).
In the
literature,  DT has been widely used as a powerful tool in computer
vision applications~\cite{fabbri}.

Another application of the DT is the computation of the medial axis
(MA) of a digital
shape~\cite{science-Rosenfeld1968},
which is defined as the set of the centers of the largest balls
contained in the treated object (see Fig. \ref{fig:DT2D3D}-$c$). MA is a very interesting representation of shape
because of its reversibility, {\it i.e.}  we can
reconstruct the original  object from its MA balls.
Many techniques have been proposed to compute the
DT and then the MA since there is a trade-off between algorithmic performance
and the \emph{accuracy} of the digital metric compared to the Euclidean one.
Hence, we can consider distances based on chamfer masks
or sequences of
chamfer distances, the
vector displacement based Euclidean distance, the Voronoi
diagram based Euclidean distance, and the square of the
Euclidean distance
 \cite{DBLP:journals/cviu/NormandSEA13}.  From a
computational point of view, several of these methods lead to time
optimal algorithms for computing the error-free Euclidean distance
transformation (EDT) for $d$-dimensional binary images
\cite{science-Maurer2003}. The extension of
these algorithms \comDav{to higher dimension} is straightforward since they use separable
techniques to compute the DT. $d$ one-dimensional operations, one per
direction of the coordinate axis, are performed.
%
Besides the efficiency of such algorithms and the various applicative areas where metric based volumetric analysis of digital objects are used, GPU implementation of such tools is still challenging.
The graphical processing unit (GPU) can be considered
  a specific parallel computing device with fine grained
  parallelism. Besides the fact that many volumetric tools are separable and rely on independent 1D  processes, each process involves distance propagations that are not well-adapted to GPU architectures.
  Recently,
  Cao proposed a banding approach that splits the 1D
  envelope computations into chunks in order to improve the parallel
  efficiency~\cite{Cao2010}. The work-load is still not optimal, but we can
  obtain a fast and error-free Euclidean DT on a GPU. Having a reliable,  efficient, and theoretically optimal algorithm for metric based volumetric processing on a GPU is still an important open problem.



\section{Homotopy, Homology, and Persistent Analysis} 

Deciding whether two objects are topologically equivalent is
important to many applications. For instance, we usually want to determine if two solid objects in 3D can be deformed to each other in 3D image recognition.  In mathematics, topological equivalence
is called homeomorphism, meaning that there is a continuous mapping between two objects and this mapping is invertible.

In general, this problem is undecidable.  However, in digital space, researchers developed many techniques for
practical data processing. Homotopic 3D thinning is one of the most successful ones.
For topological equivalence or homeomorphism, fundamental groups and homotopy groups are usually used.




\subsection{Digital Deformation and Homotopy}

Changing one object into another continuously is called
deformation. In image processing, we often see the process of morphing
one 2D or 3D picture into another. This is a type of deformation.
Computerized deformation cannot be
mathematically continuous. Computerized deformation in its digital form
 was first defined by E. Khalimsky~\cite{Che14}.
 Computerized deformation uses the digitally continuous function that is an integer function in which the value at a
digital point is the same or almost the same as its neighbors. In other words, if $x$ and
$y$ are two adjacent points in a digital space, then $| f(x)-f(y)| \le 1$~\cite{Che13}.

\begin{defi} (Homotopy~\cite{Box99})
Let $X$ and $Y$ be digital (or discrete) manifolds and
$f, g : X \rightarrow Y$ be two digital continuous functions.
$g$ is said to be digitally deformed from $f$
in $Y$ if there is a consecutive integer set $N_{m}= \{0,1,\cdots,m\}$
and a function $H: X \times N_{m}\rightarrow Y$ such that:
(1) For all $x\in X$, $H(x,0) = f (x)$ and $H(x,m) = g(x)$ ;
(2) For all $x\in X$, the induced function $H_x :N_{m}\rightarrow Y$, defined by
      $H_x =H(x,t)$ and $t\in N_{m}$, is digitally continuous; and
(3) For all $t\in N_{m}$, the induced function $H_t :X\rightarrow Y$, defined by
      $H_t =H(x,t)$ and $x\in X$,  is digitally continuous.
\end{defi}

In this definition, $f$ and $g$ are called digitally homotopic.
Now we explain the fundamental group in continuous cases: A loop  based at $x$ in a space $M$ can be viewed as a continuous function $f:[0,1]\rightarrow M$ where
$f(0)=f(1)=x$.  Fundamental groups of a space are groups made by classes
of loops (simple closed curves) that are (pointed) homotopic~\cite{Che13,Che14}.
Homotopy groups  are extensions of the fundamental
group in higher dimensions.
When $X$ is an $n$-ball, in the definition above, there is a fixed point $p$ in $Y$ where $H$
must pass through $p$.  All (pointed) homotopic mappings form a class, which is an $n$-homotopy group.
(The fundamental group and n-homotopy groups for cellular-complexes or digital spaces are not necessarily exactly the same as those in pointed space, but in most cases, they are the same.)
A direct application of the digital homotopy method is to find the skeleton of the original image.
This is called image thinning.

\subsection{Image Thinning } 

Image thinning is the extraction of the skeleton (centerline) of the image components.
To do this, the thinning algorithm usually attempts to delete as many pixels as possible while
maintaining the same homotopy groups.
For 2D images, Zhang-Suen thinning is the most practical method~\cite{Che14}.
We first introduce a 3D thinning method that was designed by Lee et al~\cite{Lee94thinning}.
The digital
surface points from Section 2 can be used to identify data points that can be deleted.
Lee et al used the Euler characteristic as the homotopic invariant
for simplicity instead of using entire homology groups.
The Euler characteristic is defined as

\begin{equation}
\chi(M) = \sum_{i \ge 0} (-1)^i K_i
\end{equation}

\noindent where $M$ is a complex and $K_i$ is the number of $i$-cells in $M$.

The key to the algorithm is deleting one or more points without changing the Euler characteristic of the remaining point set.

\begin{figure}
\begin{center}
 {\includegraphics[width=5.0cm]{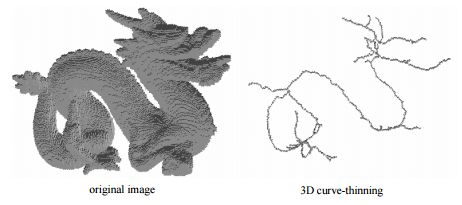}}
\end{center}
\caption{3D Thinning: $(a)$ The original 3D image, $(b)$ Its thinning result.\label{fig:thinning}}
\end{figure}

%
%
%
The Euler characteristic
can be replaced by homotopy groups for more accurate algorithms. Other thinning algorithms can also be designed. Particularly, some parallel algorithms
are designed to speed up the process~\cite{Pal12thinning}.   An example of 3D thinning is shown
in Fig. \ref{fig:thinning}~\cite{Pal12thinning}. More recent developments of 3D thinning, including comparisons of 30 different methods, can be found in ~\cite{Couprie16}.
	
Theoretical analyses of digital homotopy are hot topics in digital topology today and researchers have made much progress
~\cite{Han08,Box06}. Some discussions of homotopy and deformation can be found in~\cite{Che13}, which mainly focuses on digital functions.


\subsection{Topological Data Analysis}

In many applications, people must deal with massive datasets, especially if data points are collected by randomly arranged sensors.
This type of data is called cloud data.  Even though a discrete dataset does not mathematically have a complex topological structure,
due to the represented region size of a sampling, we could ask the following two questions: (1) Does the dataset form a special shape, and (2) Does the data
set contain a hole and how many holes are there?
The first question is related to {\it manifold learning} and the second is called {\it persistent data analysis}.

For manifold learning, our purpose is to obtain a lower dimensional manifold from a much higher dimensional space.
Persistent data analysis identifies the parameter(s) we choose in a cloud data set and presents us
with the best interpretation of the dataset. It is fortunate that in topology,
homology groups usually indicate the number of holes in each dimension. Unlike in homotopy groups, the calculation
of homology groups is relatively simple. In addition, digital homology could provide even more efficient algorithms
for important applications.

Homeomorphism, homotopy, and homology are three classical
hierarchies in understanding topological invariants. If two objects are  homeomorphic, then their homotopy groups are the same.
If two objects are homotopic, then we have the same homology (groups). Therefore, homology groups are the weaker topological
invariants compared to homeomorphism and homotopy.



\subsection{Manifold Learning and Dimension Reduction}
Manifold learning is a process used to find whether a cloud data set in a very high dimension forms a lower dimension manifold.
Two of the most popular methods for manifold learning are the Isomap and the kernel principle component analysis
~\cite{Che14}.
The idea behind Tenenbaum's Isomap method is to use the $k$-nearest neighbor ($k$-NN) or the minimum spanning tree (MST) to obtain neighborhood information. Then, we get a graph with weight
from $k$-NN and MST. Afterwards, we can use Dijkstra's algorithm to find the shortest
path between each pair of points. We present the algorithm for the Isomap below.







The kernel principle component analysis applies to situations where linear separation of the principal component is not possible~\cite{Ham2004kernel_ML}. If we perform a nonlinear transformation, called a kernel, on the original data,
then we can use the principle component analysis to obtain good results.   
In fact, reconstructing a digital manifold with a minimum total distance
to the sample points was a method suggested in ~\cite{Che14}. There are many problems we could solve in manifold learning by using digital methods.

\subsection{Persistent Homology and Data Analysis }

In recent years, persistent analysis using profound mathematics has become a powerful method in Big Data and data science~\cite{Ghr08,Car09,snavsel2017geometrical}.
For a real data set, we can
interpret a sample point in a dataset as a sample of a point, a small disk, or a cubical volume.  When the size of the disks increases,
the shape that covers the datasets may change its topological properties.
For instance, while we change the value of the radius of an assumed disk, the homology or number of holes may or may not change.
The best representation could be the homology that lasts the longest time while changing.  This method is also called
persistent homology analysis  \cite{Ghr08,Car09}.




In calculations, a more practical method involving the Vietoris-Rips complex is proposed in persistent homology analysis.
The problem of this simplex is that we need extensive calculations to construct the complex, especially when the sample set $X$ is
a large set. It requires $O(2^{|X|})$ time complexity to determine such a simplex if we directly implement a simple algorithm, as suggested by the definition  ~\cite{Che14}.
Using a digital method, we can easily solve the problem of homology groups in 2D and 3D~\cite{Che14}.


\subsection{2D Digital Holes and 3D Genus}

Hole counting, determining how many holes are in an image, is
one of the most important topological features in 2D image analysis.
There is a very simple formula to get the number of holes in 2D digital space.
Assume that
$In_p$ is the total number of corner points that direct to the inside
of the object (called inward points).  Likewise, $Out_p$ denotes the total number of corner points that direct to the outside
of the object (called outward points). We have,
\begin{equation}
          h=1+(In_p - Out_p)/4.
\end{equation}
\noindent The formal and topological proof can be found in Chapter 6 of ~\cite{Che14}. This formula was recently used in algorithm analysis for numberical analysis~\cite{matthysen2018function}. It is a very encouraging fact to digital geometry for being used in both algorithm design and computational mathematics.  

The above problem is more complex in 3D. However, we can still
get a simple formula in digital form.   For higher dimensional complexes, we may also need to use the Vietoris-Rips complex.
For a connected solid object, we obtain a simple formula for the genus ~\cite{Che14}.
This formula is the simplest form of the Gauss-Bonnet theorem in digital space.

\begin{theorem}(Chen and Rong)
Let $M$ be a closed 2D manifold in 3D space. The formula for genus is
\begin{equation}
 g = 1+ (|M_5|+2 \cdot |M_6| -|M_3|)/8.
\end{equation}
\noindent where
$M_i$ indicates the set of surface points, each of which has
$i$ adjacent points on the surface (See Fig. 3(a)).
\end{theorem}

\begin{figure}
\begin{center}
\subfigure[]{\includegraphics[width=3.cm]{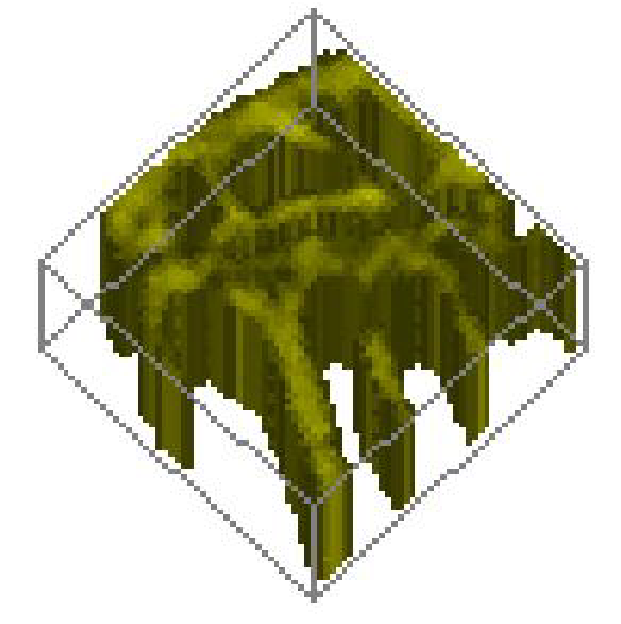}}
\subfigure[]{\includegraphics[width=4.5cm]{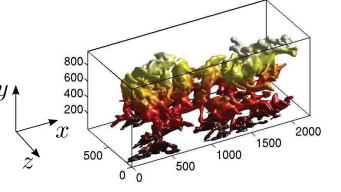}}
\end{center}
 \caption{(a) 3D genus calculation in digital space for a real bone image where $g$=10, (b) The 3D topological structure and genus \label{fig:3DHomology} }
\end{figure}

Based on this formula, we can obtain all homology groups~\cite{Che14}.
Another advantage is that the related algorithm runs in linear time.  Such computations have direct applications in medical imaging as they can be
used to identify patterns in 3D imaging, especially for bone density calculations.
The implementation of this method and other methods including digital mean curvatures to 3D image classifications
can be found in ~\cite{Che14}. A real data calculation is shown in Fig. \ref{fig:3DHomology} (a). New development on the algorithm design relating to genus of 3D objects can be found in \cite{lozano2016algorithm}; see Fig. \ref{fig:3DHomology} (b). Imiya and Eckhardt also contributed to this research ~\cite{Imiya99}.

Other application can also be addressed here, for example, calculation of $\pi$ was achieved using a method relating to digital geometry~\cite{rudolph2018recursive}. 

\section {Software Tools: Applications and Algorithms Library}

In many applications, such as material sciences or medical imaging, digital geometric tools are highly relevant for the geometric or topologic processing of shapes. Kaufman {\it et al} worked on 3D Virtual Colonoscopy , a system that is already successful in
hospital data processing~\cite{Kaufman01} (\url{http://labs.cs.sunysb.edu/labs/vislab}).
Herman led a research group who developed {\it SNARK14}: a programming system for the reconstruction of 2D images from 1D projections, which is also used in hospitals.
It contains many digital geometry methods.

DGtal (\url{http://dgtal.org})is a software system for Digital Geometry Tools and Algorithms Library
In DGtal, the library is composed of a kernel package (integers, fractions, and arithmetic properties on integers), a topological kernel (such as the ones discovered by Rosenfeld, Cartesian cellular complexes, cubical complexes, digital surface extraction, and topological invariants), a geometric package (differential estimators, fundamental objects, and volumetric processing), and digital exterior calculus packages (discrete exterior calculus operators on digital structures). We also mention an IO package that allows for easy import/export of data and visualization of structures.


\bibliographystyle{abbrv}
\bibliography{bibliography}

\end{document}

%% file: macros.tex
\newtheorem{Definition}{Definition}
\newtheorem{Theorem}{Theorem}
\newtheorem{Proposition}{Proposition}
\newtheorem{Corollary}{Corollary}
\newtheorem{Problem}{Problem}
\newtheorem{Lemma}{Lemma}
\newtheorem{Claim}{Claim}

\newcommand{\RefSection}[1]{Section~\ref{#1}}
\newcommand{\RefFigure}[1]{Fig.~\ref{#1}}
\newcommand{\RefTable}[1]{Tab.~\ref{#1}}
\newcommand{\RefDefinition}[1]{Definition~\ref{#1}}
\newcommand{\RefProposition}[1]{Proposition~\ref{#1}}
\newcommand{\RefLemma}[1]{\ensuremath{\mathrm{Lemma}~\ref{#1}}}
\newcommand{\RefTheorem}[1]{Theorem~\ref{#1}}
\newcommand{\Equ}[1]{Eq.(\ref{#1})}
\newcommand{\Implies}{\ensuremath{\Rightarrow}}
\newcommand{\st}{\ensuremath{~|~}}

\newcommand{\Z}{\ensuremath{\mathbb{Z}}}
\newcommand{\R}{\ensuremath{\mathbb{R}}}
\newcommand{\Vect}[1]{\ensuremath{\overrightarrow{#1}}}

\newcommand{\Shapes}{\ensuremath{\mathbb{X}}}
\newcommand{\Shape}{\ensuremath{X}}
\newcommand{\AnyDig}{\ensuremath{\mathtt{D}}}
\newcommand{\AnyDigF}[2]{\ensuremath{\AnyDig_{#2}(#1)}}
\newcommand{\AnyDSh}{\AnyDigF{\Shape}{h}}
\newcommand{\Dig}{\ensuremath{\mathtt{G}}}
\newcommand{\DigF}[2]{\ensuremath{\Dig_{#2}(#1)}}
\newcommand{\DSh}{\DigF{\Shape}{h}}
\newcommand{\JI}{\ensuremath{\mathtt{J}^-}}
\newcommand{\JO}{\ensuremath{\mathtt{J}^+}}
\newcommand{\JIF}[2]{\ensuremath{\JI_{#2}(#1)}}
\newcommand{\JOF}[2]{\ensuremath{\JO_{#2}(#1)}}
\newcommand{\JS}{\ensuremath{\mathtt{J}^0}}
\newcommand{\JSF}[2]{\ensuremath{\JS_{#2}(#1)}}
\newcommand{\Segment}[2]{\ensuremath{\lbrack #1 #2 \rbrack}}

\newcommand{\vz}{\ensuremath{\mathbf{z}}}
\newcommand{\vy}{\ensuremath{\mathbf{y}}}
\newcommand{\vx}{\ensuremath{\mathbf{x}}}
\newcommand{\vp}{\ensuremath{\mathbf{p}}}
\newcommand{\vu}{\ensuremath{\mathbf{u}}}
\newcommand{\vv}{\ensuremath{\mathbf{v}}}
\newcommand{\vw}{\ensuremath{\mathbf{w}}}
\newcommand{\vnu}{\ensuremath{\mathbf{\nu}}}
\newcommand{\vn}{\ensuremath{\mathbf{n}}}
\newcommand{\vZ}{\ensuremath{\mathbf{Z}}}
\newcommand{\vN}{\ensuremath{\mathbf{N}}}
\newcommand{\vc}{\ensuremath{\mathbf{c}}}
\newcommand{\NU}{\ensuremath{\mathcal{U}}}
\newcommand{\Id}{\ensuremath{\mathrm{Id}}}
\newcommand{\SO}{\ensuremath{\mathrm{S}}}

\newcommand{\Body}[2]{\ensuremath{\lbrack #1 \rbrack_{#2}}}

\newcommand{\Mom}[1]{\ensuremath{m^{#1}}}
\newcommand{\DMom}[2]{\ensuremath{\hat{m}_{#2}^{#1}}}
\newcommand{\Cov}[0]{\ensuremath{\mathcal{V}}}
\newcommand{\DCov}[1]{\ensuremath{\Cov_{#1}}}
\newcommand{\Q}[1]{\ensuremath{Q({#1})}}
\newcommand{\hQ}[2]{\ensuremath{Q_{#2}({#1})}}

\newcommand{\Bd}[1]{\ensuremath{\partial #1}}
\newcommand{\Int}[1]{\ensuremath{\mathrm{Int}(#1)}}
\newcommand{\SD}{\ensuremath{\Delta}}
\newcommand{\dS}{\Bd{X}}
\newcommand{\dD}{\Bd{\DSh}}

\newcommand{\Proj}{\pi_h^\Shape}
\newcommand{\Card}{\mathrm{Card}}

\newcommand{\q}{\ensuremath{\kappa}}
\newcommand{\estq}[1]{\ensuremath{\hat{\kappa}_{#1}}}
\newcommand{\speed}[1]{\ensuremath{\tau_{#1}}}
\newcommand{\eabs}{\ensuremath{\epsilon_{abs}}}
\newcommand{\erel}{\ensuremath{\epsilon_{rel}}}
\newcommand{\eavgabs}{\ensuremath{\overline{\epsilon}_{abs}}}
\newcommand{\eavgrel}{\ensuremath{\overline{\epsilon}_{rel}}}

\newcommand{\com}[1]{\textcolor{red}{\it #1}}
\newcommand{\comDav}[1]{\textcolor{blue}{ #1}}
\newcommand{\comJeremy}[1]{\textcolor[rgb]{0.26,0.73,0.06}{\it #1}}
\newcommand{\comJaco}[1]{\textcolor{magenta}{#1}}

\newcommand{\Area}{\ensuremath{\mathrm{Area}}}
\newcommand{\Vol}{\ensuremath{\mathrm{Vol}}}
\newcommand{\AreaC}[0]{\ensuremath{\widehat{\Area}}}
\newcommand{\VolC}[0]{\ensuremath{\widehat{\Vol}}}
\newcommand{\EqDef}{\ensuremath{:=}}
\newcommand{\Curv}{\ensuremath{\kappa}}
\newcommand{\FirstCurv}{\ensuremath{\kappa_1}}
\newcommand{\SecondCurv}{\ensuremath{\kappa_2}}
\newcommand{\FirstCurvDir}{\ensuremath{\mathbf{w}_1}}
\newcommand{\SecondCurvDir}{\ensuremath{\mathbf{w}_2}}
\newcommand{\NormalDir}{\ensuremath{\mathbf{n}}}
\newcommand{\MeanCurv}{\ensuremath{H}}
\newcommand{\CurvT}[1]{\ensuremath{\tilde{\Curv}^{#1}}}
\newcommand{\MeanCurvT}[1]{\ensuremath{\tilde{\MeanCurv}^{#1}}}
\newcommand{\PCurvHat}[1]{\ensuremath{\hat{\Curv}^{#1}}}
\newcommand{\DCurvHat}[1]{\ensuremath{\hat{\Curv}_D^{#1}}}
\newcommand{\PMeanCurvHat}[1]{\ensuremath{\hat{\MeanCurv}^{#1}}}
\newcommand{\PFirstCurvHat}[1]{\ensuremath{\hat{\Curv}_1^{#1}}}
\newcommand{\PSecondCurvHat}[1]{\ensuremath{\hat{\Curv}_2^{#1}}}
\newcommand{\PFirstCurvDirHat}[1]{\ensuremath{\hat{\mathbf{w}}_1^{#1}}}
\newcommand{\PSecondCurvDirHat}[1]{\ensuremath{\hat{\mathbf{w}}_2^{#1}}}
\newcommand{\PNormalDirHat}[1]{\ensuremath{\hat{\mathbf{n}}^{#1}}}
\newcommand{\PFirstCurvDir}[1]{\ensuremath{\tilde{\mathbf{e}}_1^{#1}}}
\newcommand{\PSecondCurvDir}[1]{\ensuremath{\tilde{\mathbf{e}}_2^{#1}}}
\newcommand{\PNormalDir}[1]{\ensuremath{\tilde{\mathbf{e}}_3^{#1}}}

\newcommand{\DFr}[1]{\ensuremath{\Delta_{#1}}}
\newcommand{\hBd}[1]{\ensuremath{\partial_{#1}}}

\newcommand{\Ball}[2]{\ensuremath{B_{#1}(#2)}}
\newcommand{\HalfBall}[2]{\ensuremath{B^+_{#1}(#2)}}
\newcommand{\Rounded}[2]{\ensuremath{\left\lbrack \frac{#1}{#2} \right\rbrack}}

\newcommand{\MA}[1]{\ensuremath{\mathrm{MA}(#1)}}
\newcommand{\Reach}[1]{\ensuremath{\mathrm{reach}(#1)}}
\newcommand{\ProjX}[1]{\ensuremath{\pi^{#1}}}
\newcommand{\Dist}{\ensuremath{\mathfrak{d}}}

\newcommand{\BdZ}[1]{\ensuremath{Bd({#1})}}

\newcommand{\PCurvHatStar}{\ensuremath{\hat{\Curv}^{*}}}
\newcommand{\PMeanCurvHatStar}{\ensuremath{\hat{H}^{*}}}